\documentclass[aps,prb,twoside,twocolumn,showpacs,floatfix,superscriptaddress,citeautoscript]{revtex4-1}

\usepackage{graphicx}
\usepackage{amsmath}
\usepackage{units}
\usepackage{hyperref}
 
\newcommand{\etal}[0]{\textit{et al.}}

\newcommand{\eV}[0]{\text{eV}}
\newcommand{\meV}[0]{\text{meV}}
\newcommand{\AAA}[0]{\text{\AA}}

\newcommand{\sect}[1]{Sect.~\ref{#1}}
\newcommand{\fig}[1]{Fig.~\ref{#1}}
\newcommand{\eq}[1]{Eq.~\eqref{#1}}

\renewcommand{\vec}[1]{\ensuremath\boldsymbol{#1}}
\renewcommand{\epsilon}[0]{\varepsilon} 
\newcommand{\braket}[3]{\bigl{<}#1\big|#2\big|#3\bigr{>}}

\newcommand{\ket}[1]{\big|#1\bigr{>}}
\newcommand{\psih}[0]{\widehat{\psi}}
\newcommand{\psihd}[0]{\widehat{\psi}^{\dagger}}
\newcommand{\ibraket}[2]{\bigl{<}#1\big|#2\bigr{>}}
\newcommand{\rb}[0]{{\boldsymbol r}}
\newcommand{\Rb}[0]{{\boldsymbol R}}
\newcommand{\kb}[0]{{\boldsymbol k}}

\begin{document}
 

\title{
  Quasi-particle spectra, absorption spectra, and excitonic properties \\
  of sodium iodide and strontium iodide from many-body perturbation theory
}

\author{Paul Erhart}
\affiliation{
  Chalmers University of Technology, Department of Applied Physics, Gothenburg, Sweden
}
\affiliation{
  Physical and Life Sciences Directorate, Lawrence Livermore National Laboratory, Livermore, California 94550, USA
}
\author{Andr{\'e} Schleife}
\author{Babak Sadigh}
\author{Daniel {\AA}berg}
\affiliation{
  Physical and Life Sciences Directorate, Lawrence Livermore National Laboratory, Livermore, California 94550, USA
}

\begin{abstract}
We investigate the basic quantum mechanical processes behind non-proportional response of scintillators to incident radiation responsible for reduced resolution. For this purpose, we conduct a comparative first principles study of quasiparticle spectra on the basis of the $G_0W_0$ approximation as well as absorption spectra and excitonic properties by solving the Bethe-Salpeter equation for two important systems, NaI and SrI$_2$. The former is a standard scintillator material with well-documented non-proportionality while the latter has recently been found to exhibit a very proportional response. We predict band gaps for NaI and SrI$_2$ of 5.5 and 5.2\,eV, respectively, in good agreement with experiment. Furthermore, we obtain binding energies for the groundstate excitons of 216\,meV for NaI and $195\pm25\,\meV$ for SrI$_2$. We analyze the degree of exciton anisotropy and spatial extent by means of a coarse-grained electron-hole pair-correlation function. Thereby, it is shown that the excitons in NaI differ strongly from those in SrI$_2$ in terms of structure and symmetry, even if their binding energies are similar. Furthermore, we show that quite unexpectedly the spatial extents of the highly anisotropic low-energy excitons in SrI$_2$ in fact exceed those in NaI by a factor of two to three in terms of the full width at half maxima of the electron-hole pair-correlation function.
\end{abstract}

\maketitle

\section{Introduction}

The ability to detect high-energy radiation is crucial for applications in high-energy physics, medicine, as well as homeland security \cite{Rod97, Kno10}. For these applications, one frequently relies on scintillators, which are materials that convert high energy radiation into low energy photons. By correlating the number of photons generated during a short time interval with the energy of an incident radiation quantum, it is possible to deduce the energy spectrum of the incoming signal. Provided the energy resolution is sufficiently high this in principle allows identification of the radiation source. The energy resolution that is achievable using classical scintillators such as NaI and CsI is, however, limited and does not meet the requirements for radioactive isotope identification \cite{NelGosKna11}.

An analysis of counting statistics demonstrates that resolution improves with an increase in luminosity, which usually results from a higher conversion efficiency, i.e. relatively more photons are generated per incident energy. Yet the resolution of most scintillators is significantly worse than what can be expected based on their respective luminosities \cite{Dor10}. During the last decades it has been established that this discrepancy can be traced to the non-linear response of the material to the energy of the incident radiation quantum, which accordingly has been studied intensively \cite{DorHaaEij95, RooVal96, Vas08, MosPayCho08, MosBizWil12}.

While the fundamental physical mechanisms from track creation to final photon emission are understood at least on a qualitative level, the relative importance of each of these events is still unclear. In spite of several recent promising attempts to resolve this situation \cite{PayCheHul09, KerRosCan09, BizMosSin09, WilGriLi11, WanWilGri13}, there is currently no established numerical framework that has successfully combined the aforementioned physical mechanisms into a predictive model. One of the major reasons is the enormous complexity and uncertainty connected with thermalization and transport of excitation carriers (primarily electron, holes, and excitons) as well as their respective contributions to the response of the system.

Existing models \cite{PayCheHul09, KerRosCan09, BizMosSin09, WilGriLi11, WanWilGri13} typically rely on a number of physical parameters, such as dielectric function, migration barriers of self-trapped excitons, electron and hole mobilities, defect trapping rates, as well as Auger recombination rates of free carriers and excitonic states. Since at least some of these quantities are notoriously difficult to measure experimentally, parameter-free electronic-structure calculations are very valuable for providing not only physical insight but input data for such models. They also serve to complement and guide experimental efforts. The impact of electronic properties such as fundamental band gaps and charge carrier mobilities has recently been studied for a number of scintillating materials \cite{SetGauRom09, LiGriWil11}. There remains, however, a pronounced gap in our knowledge and understanding of both quantities and processes that relate to the interaction of charge carriers and excitations, specifically excitonic effects. This is at least in part because the methods required to access these effects come with a significant computational burden.

Two materials that are known to be very promising for scintillator applications are LaBr$_3$:Ce (see e.g., Ref.~\onlinecite{LoeDorEij01}) and SrI$_2$:Eu (see e.g., Refs.~\onlinecite{CheHulDro08, WilLoeGlo08, HawGroCui08, ChePayAsz09}), both of which exhibit very high luminosity and significantly improved energy resolution compared to NaI. While LaBr$_3$ has already been characterized rather extensively both experimentally \cite{DorLoeVin06, DotMcGHar07, Dor10, BizDor07} and theoretically \cite{BizDor07, VanJafKer10, CanChaBou11, AndKolDor07, Sin10, McIGaoTho07, AbeSadErh12}, information on the properties of SrI$_2$, in particular its electronic structure, is sparse \cite{Sin08}. Even fundamental quantities such as the band gap are still to be determined consistently and the influence of excitonic and polaronic effects is entirely unknown at this point. Yet, a deeper understanding of this material is crucial since properties such as the band gap, exciton binding energies and dielectric functions constitute essential building blocks for any theoretical study of e.g., free carrier transport, polaron formation and migration as well as Auger recombination. In addition, they represent important parameters for the interpretation of experimental data.

The objective of the present work is to alleviate this situation by providing predictions for electronic and optical properties on the basis of modern theoretical spectroscopy techniques. We specifically aim at accurately describing band structures and densities of states (including fundamental band gaps), optical absorption spectra, exciton binding energies and exciton localization. In order to assess the reliability of our parameter-free simulations and to put our results in context, we also carry out calculations for the classic scintillator NaI, for which extensive experimental data are available.

The remainder of this paper is organized as follows. In \sect{sect:computational_parameters} we summarize the computational approach employed in this work. A convenient decomposition of the six-dimensional exciton wave function into single-particle densities and an envelope function is introduced in \sect{sect:envelope_function}. Results for band structure, dielectric properties and absorption spectra are presented in Sects.~\ref{sect:qp} and \ref{sect:absorption}. Exciton binding energies as well as an analysis of their spatial extent are the subject of Sects.~\ref{sect:binding_energies} and \ref{sect:excdens}. Finally, we summarize the results and discuss them in the context of the scintillation properties of NaI and SrI$_2$ in \sect{sect:discussion}.

\section{Theoretical approach}

Since this work aims at a precise description of electronic structure as well as optical properties it requires techniques beyond standard density functional theory (DFT). Specifically, it is essential to employ computational schemes that are capable of describing quasiparticle (QP) and excitonic effects. To this end, we use a combination of DFT and $G_0W_0$ calculations \cite{Hed65} to describe single-particle excitations that govern band structures and densities of states. Two-particle (electron-hole) excitations have to be taken into account when computing exciton binding energies, dielectric functions, and optical absorption spectra. This is achieved by solving the Bethe-Salpeter equation (BSE) for the optical polarization function \cite{SalBet51, OniReiRub02}.

\subsection{Computational parameters}
\label{sect:computational_parameters}

In this work we use experimental values for the crystallographic geometries. Sodium iodide adopts the rocksalt structure with a lattice constant of 6.48\,\AA. SrI$_2$ belongs to space-group Pbca (number 61 in the International Tables of Crystallography, Ref.~\onlinecite{itca}). Its unit cell contains 24 atoms with nine internal degrees of freedom. The experimentally determined lattice parameters are $a=15.22\,\AAA$, $b=8.22\,\AAA$, and $c=7.90\,\AAA$ with the following internal coordinates: Sr on Wyckoff site $8c$ ($x=0.1105$, $y=0.4505$, $z=0.2764$), I(1) on Wyckoff site $8c$ ($x=0.2020$, $y=0.1077$, $z=0.1630$), and I(2) on Wyckoff site $8c$ ($x=-0.0341$, $y=0.2682$, $z=0.0054$) \cite{BarBecGru69}.

All calculations were carried out using the projector-augmented wave method to describe the electron-ion interaction \cite{Blo94, KreJou99}. We used a plane-wave expansion for the wave functions with a cutoff energy of 228\,eV for both materials. DFT and $G_0W_0$ electronic structures were generated using the Vienna \emph{Ab-initio} Simulation Package \cite{KreHaf93, KreHaf94, KreFur96a, KreFur96b, ShiKre06}. The corresponding BSE implementation has been discussed in Refs.~\onlinecite{SchGluHah03, FucRodSch08, RodFucFur08}. 

For NaI and SrI$_2$ we use the generalized-gradient approximation \cite{PerBurErn96} and the local-density approximation \cite{CepAld80}, respectively, to represent exchange-correlation effects at the DFT level. Brillouin-zone (BZ) integrations for both DFT and $G_0W_0$ calculations were carried out by summing over $\Gamma$-centered $6\,\times\,6\,\times\,6$ Monkhorst-Pack\cite{MonPac76} (MP) grids in the case of NaI. For SrI$_2$ the density of states (DOS) was computed on the DFT level using a $6\,\times\,11\,\times\,11$ $\Gamma$-centered MP $\kb$-point grid while the band structure was obtained on the basis of a $4\,\times\,7\,\times\,7$ mesh. Calculations of $G_0W_0$ QP energies were carried out for several $\kb$-point grids up to $\Gamma$-centered $3\,\times\,4\,\times\,4$ and up to 2880 bands were included in the calculations to achieve convergence of the dielectric function entering the screened interaction $W$. Spin-orbit coupling (SOC) was taken into account using the projector-augmented wave implementation described in Ref.~\onlinecite{AbeSadErh12}. Based on convergence tests we estimate that these computational parameters yield QP shifts around the band edges that are converged to within 50\,meV for both materials.

In order to describe optical properties, we calculated dielectric functions using regular MP meshes of $16\,\times\,16\,\times\,16$ and $4\,\times\,6\,\times\,6$ $\kb$-points for NaI and SrI$_2$, respectively. For a more efficient sampling of the BZ, each grid was displaced by a small random vector. For NaI and SrI$_2$ we used 32 and 48 conduction bands, respectively. In addition, the number of Kohn-Sham states contributing to the BSE Hamiltonian is limited by the BSE cutoff energy which specifies the maximum non-interacting electron-hole pair energy that is taken into account. Here we used a BSE cutoff energy of at least 13.0 and 5.0\,eV for NaI and SrI$_2$ to set up the excitonic Hamiltonian from independent electron-hole pairs. The screened Coulomb interaction $W$ was constructed assuming the $\boldsymbol q$-diagonal model function of Bechstedt \etal\ \cite{BecDelCap92} and static electronic dielectric constants of 3.69 and 4.58 as obtained for NaI and SrI$_2$ on the DFT level.

The Coulomb singularity present in the BSE Hamiltonian effectively prevents accurate one-shot calculations of exciton binding energies for a given $\kb$-point mesh \cite{PusAmb02, FucRodSch08}. This is somewhat alleviated by the introduction of so-called singularity corrections. Still present implementations are left with error terms that are proportional to the inverse number of $\kb$-points. Therefore, the most efficient scheme currently consists of extrapolating calculated binding energies for a number of $\kb$-point grids to the continuum limit \cite{FucRodSch08}. We sample the BZ using both regular and hybrid $\kb$-point meshes as defined in Ref.~\onlinecite{FucRodSch08}. For NaI we employed regular $\Gamma$-centered grids up to $21\times21\times21$ as well as $11^3:6^3:x^3$ hybrid grids with $x$=$\{22, 25\,\nicefrac[]{2}{3}, 29\,\nicefrac[]{1}{3}, 33\}$. For SrI$_2$ the finest regular meshes we used were $7\times 9\times 9$, $5\times 11\times 11$, and $3\times 13\times 13$ and, in addition, we employed hybrid meshes of $5\times5\times5:4\times2\times2:x$ with $x$=$\{5\times15\times15,5\times20\times20, 5\times25\times25 \}$ for finer sampling around the $\Gamma$-point in the $\kb_y$ and $\kb_z$ directions.

\subsection{Electron-hole densities}
\label{sect:envelope_function}

The electron-hole separation of a free exciton is readily obtained within the effective mass approximation for a two-band model \cite{Ell57}. In cases where this approximation is not applicable, the BSE wave functions must be analyzed instead. To this end, Dvorak \etal\ have computed the relative distribution of the electron around the hole as well as spread and localization lengths by a straightforward integration of the full excitonic wave function \cite{DvoWeiWu13}. The BSE wave functions and electron hole distribution function are, however, \mbox{six-dimensional} functions of the electron and hole coordinates which renders the problem, at least computationally, more involved. Furthermore, when discussing the spatial distribution of electrons or holes, a charge density (or wave function) is usually constructed by fixing the electron (hole) at the position of an atom belonging to the conduction (valence) band edge \cite{RohLou98, IsmLou05, HumPusAmb04, CudAttTok10}. This choice is though somewhat arbitrary since the Bloch states may be strongly hybridized. To alleviate this problem and to avoid the representation of a density over the entire Born-von-K\'arm\'an cell on a dense Cartesian mesh in the case of calculating the electron-hole separation, we follow another route.

The electron-hole pair distribution function $\rho(\rb_e, \rb_h)$ specifies the probability of simultaneously finding an electron at $\rb_e$ and a hole at $\rb_h$. In the single-particle band picture we thus can write
\begin{align}
  \rho(\rb_e, \rb_h)=\rho_e(\rb_e) \rho_h(\rb_h).
\end{align}
In this approximation both electron and hole can be completely delocalized over the entire crystal. However, since electron and hole are coupled through the Coulomb interaction the distribution function assumes a more complex form
\begin{align}
  \rho(\rb_e, \rb_h)=\rho_e(\rb_e) \rho_h(\rb_h) g_{eh}(\rb_e, \rb_h),
\end{align}
where $g_{eh}$ represents the explicit pair correlation between electron and hole. Thus, given an electron at $\rb_e$ the probability of finding a hole at $\rb_h$ is $\rho_h(\rb_h) g_{eh}(\rb_e, \rb_h)$. In the following we will derive explicit expressions for the excitonic single-particle densities and an approximate correlation function that turns out to be solely a function of the electron-hole separation. Thereby, we will obtain a partitioning of the pair distribution function into two single-particle {\em cell-periodic} densities, representing the local variations of electron and hole densities, as well as an associated pair distribution function that is coarse-grained over each periodic cell.

The exciton wave function can be represented in terms of electron and hole coordinates via the amplitude \cite{Str88, RohLou00}
\begin{align}
  \chi_j(\rb_e, \rb_h) &= \braket{N;0}{\psihd(\rb_e)\psih(\rb_h) }{N;j},
\end{align}
where $\ket{N;j}$ is the $j$-th excited state of the $N$-electron system, and $\psih(\rb)$ and $\psihd(\rb)$ are the standard annihilation and creation operators acting on coordinate $\rb$, respectively. Translated into the basis of Bloch orbitals this becomes
\begin{align}
  \chi_j(\rb_e, \rb_h) &= \sum_{\kb c v} A^j_{\kb c v} \phi^*_{\kb v}(\rb_h) \phi_{\kb c}(\rb_e),
  \label{eq:exc_wf}
\end{align}
where the coefficients $A^j_{\kb c v}$ are the eigenvectors of the BSE matrix. Therefore we can write the associated electron-hole density as
\begin{align}
  \rho^{j}_{eh}(\rb_e, \rb_h) &=  \sum_{\kb c v}  \sum_{\kb' c' v'} A^{j*}_{\kb c v} A^{j}_{\kb' c' v'} \notag \\
  & \times \phi_{\kb v}(\rb_h) \phi^*_{\kb c}(\rb_e) \phi^*_{\kb' v'}(\rb_h) \phi_{\kb' c'}(\rb_e). \label{eq:excden}
\end{align}

By integrating over either the electron or hole coordinate we obtain the cell-periodic hole and electron charge densities, respectively,
\begin{subequations}
  \begin{align}
    \rho^j_e(\rb_e) &= \sum_\kb \sum_{c\,c'} \phi^*_{\kb c}(\rb_e) \phi_{\kb' c'}(\rb_e) \sum_v A^{j*}_{\kb c v} A^{j}_{\kb c' v}, \label{eq:rhoe} \\
    \rho^j_h(\rb_h) &= \sum_\kb \sum_{v\,v'}  \phi_{\kb v}(\rb_h) \phi^*_{\kb' v'}(\rb_h) \sum_c A^{j*}_{\kb c v} A^{j}_{\kb c v'}. \label{eq:rhoh}
  \end{align} 
\end{subequations}
We note that the electron-hole density of the entire $N$-electron system can be written as $\rho^j=\rho^{0}+\rho^j_e-\rho^j_h$, where $\rho^0$ is the ground state charge density. This can be seen by expanding the $N$-electron exciton wave function $\chi_j(\rb_1,\ldots,\rb_N)=\ibraket{\rb_1,\ldots,\rb_N}{N;j}$ into Slater determinants.

Returning to the question of electron-hole separation we again consider the two-particle density of \eq{eq:excden} and make the variable substitution $\rb_{e/h}=\rb'_{e/h}+\Rb_{e/h}$, where now $\rb'_{e/h}$ is constrained to one unit cell and $\Rb_{e/h}$ denotes a lattice vector. By integrating the resulting two-particle density over $\rb'_e$ and $\rb'_h$ we obtain a coarse-grained pair distribution function
\begin{align}
  \widetilde{g}^{j}_{eh}(\Rb_e, \Rb_h) &=\sum_{\kb c v}  \sum_{\kb' c' v'} A^{j*}_{\kb c v} A^{j}_{\kb' c' v'} \nonumber \\
  \times & \exp \left[ 
    i\left(\kb'-\kb\right)\cdot \left(\Rb_e-\Rb_h\right)
    \right] I^*_{\kb v,\kb' v'} I_{\kb c,\kb' c'}, \label{eq:envelope}
\end{align}
where $I_{\kb n,\kb' n'}=\int d\rb \phi^*_{\kb n}(\rb) \phi_{\kb' n'}(\rb)$ is an integral over one unit cell only. Also note that $\widetilde{g}_{eh}$ of \eq{eq:envelope} is only a function of the electron-hole separation $\Rb_e-\Rb_h$ and thus resembles an envelope function. As a result, we can approximately write the electron-hole pair distribution function as
\begin{align}
\rho^j_{eh}(\rb_e+\Rb_r,\rb_h+\Rb_h) =
  \rho^j_e(\rb_e)\rho^j_h(\rb_h) \widetilde{g}^{j}_{eh}(\Rb_e-\Rb_h).
  \label{eq:eh_pair_distribution}
\end{align}
In this simplified picture, we can therefore write the exciton pair distribution function as an envelope function $\widetilde{g}_{eh}$ modulated by the product of two periodic single-particle charge densities, $\rho^j_e$ and $\rho^j_h$.

\section{Results}

\subsection{QP band structures}
\label{sect:qp}

\newcommand{\myscale}{0.65}
\begin{figure}
\includegraphics[scale=\myscale]{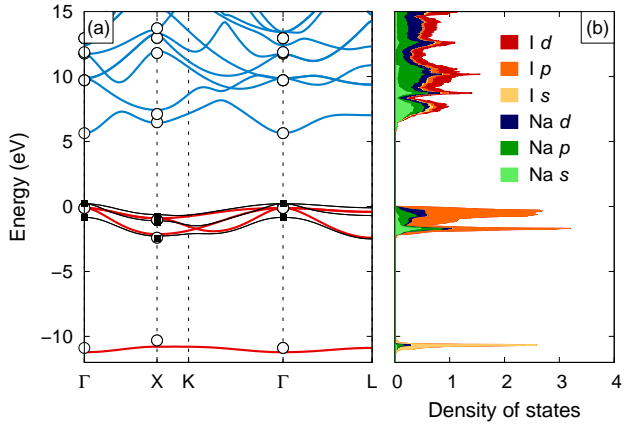}
  \caption{
    (a) NaI non-relativistic QP energies from $G_0W_0$ (empty circles) superimposed on DFT+$\Delta$ band structure (colored lines). The spin-orbit split valence bands are shown by black lines and filled squares. (b) Partial density of states for NaI as obtained from DFT+$\Delta$ calculations.
  }
  \label{fig:nai_bands}
\end{figure}

\begin{figure*}
\includegraphics[scale=\myscale]{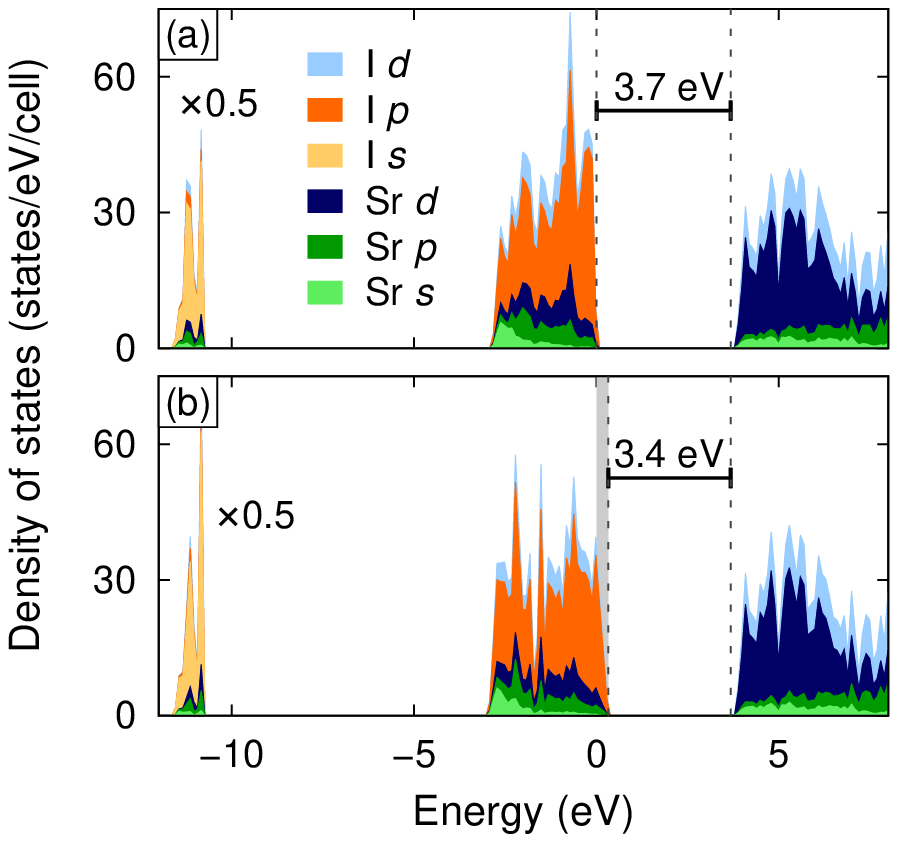}
\includegraphics[scale=\myscale]{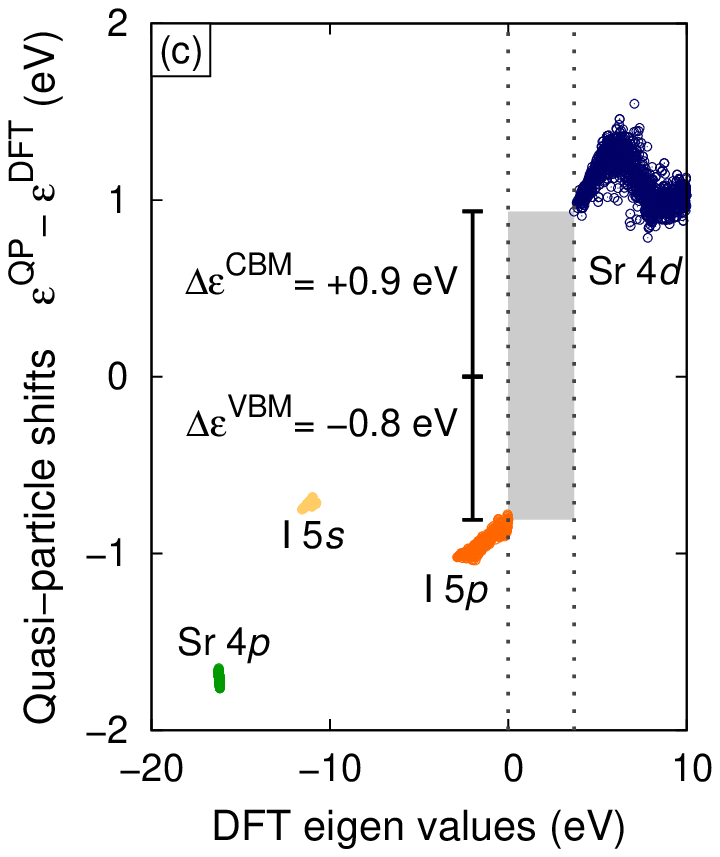}
\includegraphics[scale=\myscale]{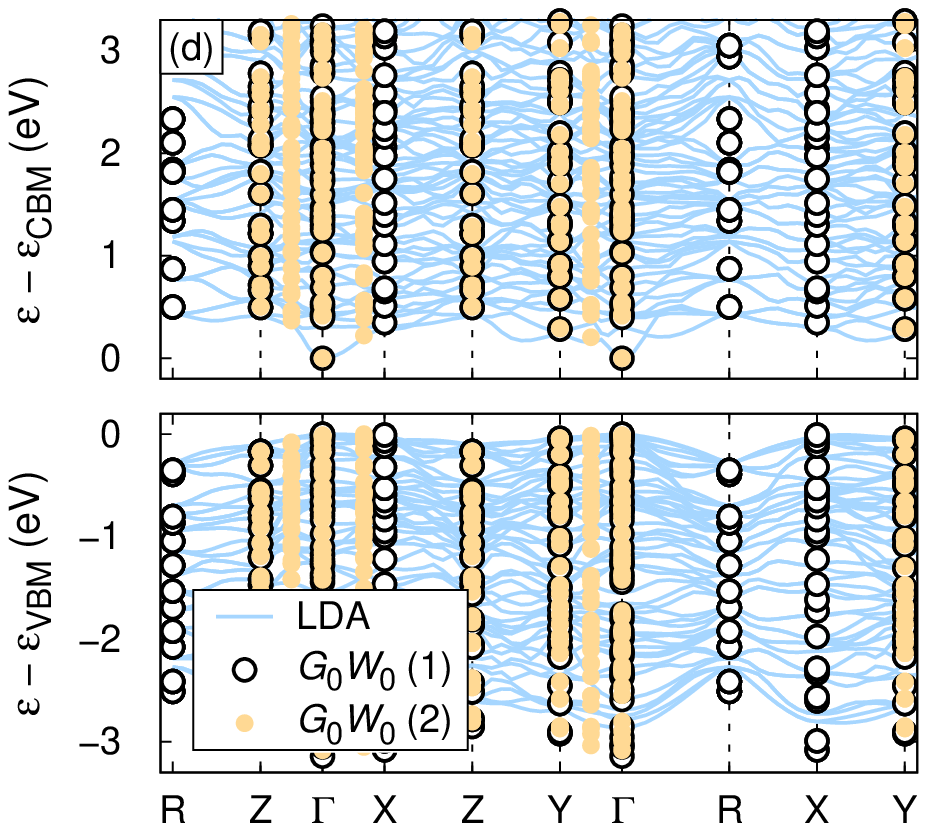}
  \caption{
    Partial density of states for SrI$_2$ obtained from DFT calculations (a) without and (b) with spin-orbit coupling. The major change upon the inclusion of SOC effects is an upward shift of the valence band edge by 0.3\,eV, which is indicated by the gray bar in panel (b).
    (c) QP energy shifts from $G_0W_0$ calculations with respect to initial DFT eigenenergies.
    (d) QP energies from $G_0W_0$ superimposed onto DFT+$\Delta$ band structure. The two different sets of $G_0W_0$ QP energies correspond to $\Gamma$-centered $\kb$-point grids with (1) $2\times2\times2$ and (2) $3\times4\times4$ divisions, respectively.
  } 
  \label{fig:SrI2_dos}
  \label{fig:SrI2_gw}
\end{figure*}

In the case of NaI DFT calculations yield a band gap of 3.7\,eV. This value reduces to 3.4\,eV when SOC is taken into account, which is considerably less than experimental values that lie in the range between 5.8 and 6.3\,eV \cite{TeeBal67, BroGahFuj70, PooJenLec74}. The $G_0W_0$ band gap is 5.8\,eV without and 5.4\,eV with SOC in much better agreement with experiment, see \fig{fig:nai_bands}. The major effect of SOC is to cause a splitting of the I\,$5p$ band by about 0.9\,eV at the $\Gamma$-point.

The partial DOS of SrI$_2$ is shown in \fig{fig:SrI2_dos} calculated both without and with SOC taken into account. The uppermost valence band is composed almost exclusively of I\,$5p$ states whereas the conduction band is dominated by Sr\,$4d$ states with a contribution from Sr\,$5s$ states at the bottom of the conduction band. The largest change due to SOC is a splitting of 1.15\,eV observed for the Sr\,$4p$ states that lie between $-15.8$ and $-17.0\,\eV$ below the valence band maximum. The band gap decreases by 0.3\,eV upon inclusion of SOC due to a shift of the valence band maximum (VBM). Otherwise the structure of both valence and lower conduction band states is largely preserved. Figure~\ref{fig:SrI2_gw}(c) demonstrates that there are no qualitative changes in going from DFT to $G_0W_0$ within a band manifold. The valence band width increases from 2.86\,eV to 3.00\,eV and the band gap from 3.7 to 5.5\,eV. The band characters and their ordering are only weakly affected and, as a result, a rigid upward shift (a scissor correction, referred to as DFT+$\Delta$ from here on) of the DFT conduction band yields reasonable agreement between DFT and DFT+$G_0W_0$ QP energies. This is demonstrated in \fig{fig:SrI2_gw}(d), which shows QP energies from $G_0W_0$ superimposed on a DFT band structure.
  
Due to the prohibitive computational cost, SOC effects were not taken into account on the $G_0W_0$ level for SrI$_2$. Our calculations for NaI as well as LaBr$_3$ (see Ref.~\onlinecite{AbeSadErh12}), however, show that DFT and $GW$ yield similar results for SOC induced shifts. We therefore can employ the band gap reduction calculated on the DFT level [compare \fig{fig:SrI2_dos}(a,b)] to estimate the SOC corrected band gap of SrI$_2$ as 5.2\,eV.

\subsection{Dielectric functions and absorption spectra}
\label{sect:absorption}

\begin{figure}
\includegraphics[scale=\myscale]{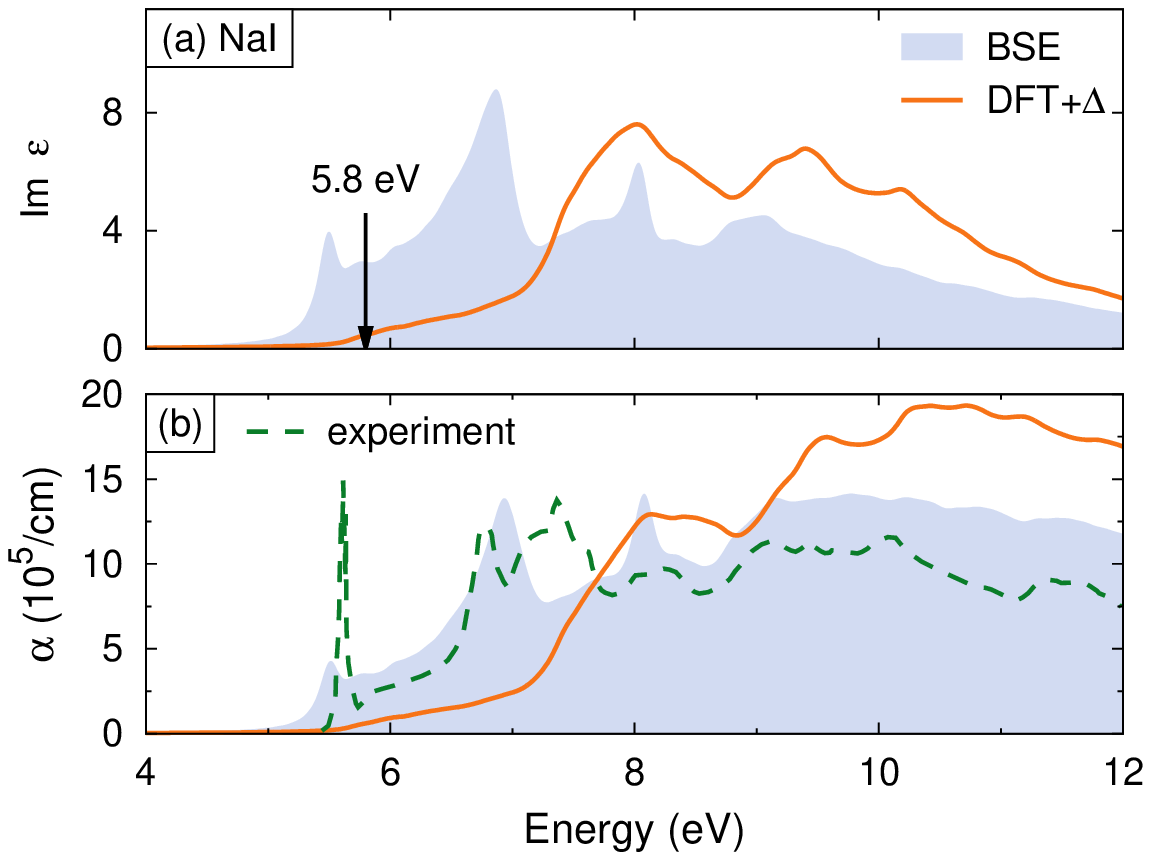}
  \caption{
    (a) Imaginary part of the dielectric function and (b) absorption spectrum for NaI. The vertical arrow indicates the fundamental QP band gap (excluding SOC effects). The experimental absorption data is taken from Ref.~\onlinecite{TeeBal67}.
  }
 \label{fig:dielec}
\end{figure}

\begin{figure}
\includegraphics[scale=\myscale]{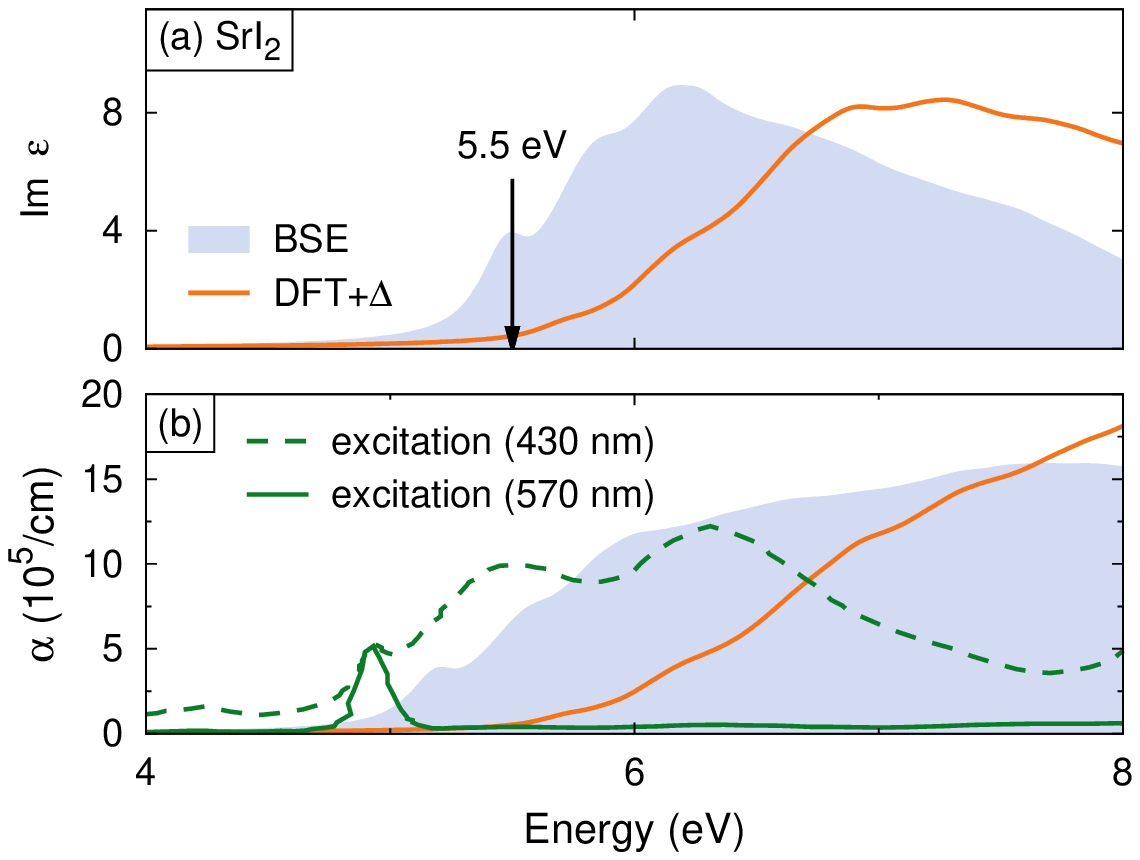}
  \caption{
    (a) Imaginary part of the dielectric function and (b) absorption spectrum for SrI$_2$. Panel (b) also shows excitation spectra recorded by Pankratov \etal\ (Ref. \onlinecite{PanPopShi13}). The vertical arrow indicates the fundamental QP band gap (excluding SOC effects).
  }
  \label{fig:absorp}
\end{figure}

We now address two-particle excitations and optical properties based on the dielectric function. As discussed above, the good agreement between DFT+$G_0W_0$ and rigidly shifted DFT band structures (DFT+$\Delta$) enables us to use the latter as the starting point for the BSE calculation. In this work we use values of 2.03\,eV and 1.82\,eV for the rigid shift $\Delta$ for NaI and SrI$_2$, respectively.

We first study the spectroscopic properties of NaI by comparing the imaginary part of the dielectric function computed using DFT+$\Delta$ to the one obtained from BSE calculations. Figure~\ref{fig:dielec} reveals a pronounced peak with large oscillator strength near the absorption onset related to an excitonic bound state, which is not captured by the single-particle DFT picture approximation. Overall the BSE result exhibits a pronounced redistribution of peak weights leading to structural changes in the dielectric function. One also notices a red shift of the entire spectrum in going from DFT+$\Delta$ to BSE. These features are attributed to excitonic effects and are also apparent in the predicted absorption spectrum, which is shown in \fig{fig:absorp}(b) along with experimental data recorded at 10\,K \cite{TeeBal67}. The BSE spectrum is overall in good agreement with the experimental data, in particular if compared to the DFT+$\Delta$ result. The most pronounced difference occurs between 6.5 and 7.5\,eV where the experimental spectrum exhibits two peaks whereas there is only one distinct feature in the BSE data.

On the basis of the similarity between the absorption spectra of gaseous xenon and the iodide ion Teegarden and Baldini \cite{TeeBal67} proposed that the two lowest peaks, which are separated by approximately 1\,eV, originate from the spin-orbit split $5p6s$ atomic state. In fact the separation of the two features is comparable to the spin-orbit splitting of 0.9\,eV calculated for the I\,$4p$ band on a DFT/$G_0W_0$ level [see \fig{fig:SrI2_dos}(a)]. Unfortunately SOC is currently not included in our BSE calculations and thus we cannot directly assess this assignment. It should, however, be noted that the separation of the second and third peak in the experimental spectrum, which overlap with a single peak in the BSE spectrum, also exhibit a separation of about 0.9\,eV.

Figure~\ref{fig:dielec} displays the predicted dielectric function as well as absorption and transmission spectra at the DFT+$\Delta$ and BSE levels. As in the case of NaI the BSE spectra show a strong red shift with respect to DFT+$\Delta$ and a redistribution of spectral weight due to excitonic effects. The absorption spectra are rather featureless but clearly the onset of absorption is below the fundamental QP band gap and a shoulder is apparent around 5.5\,eV. The latter feature is corroborated by experimental emission spectra \cite{PanPopShi13}.

\subsection{Exciton binding energies}
\label{sect:binding_energies}

\begin{figure}
  \centering
\includegraphics[scale=\myscale]{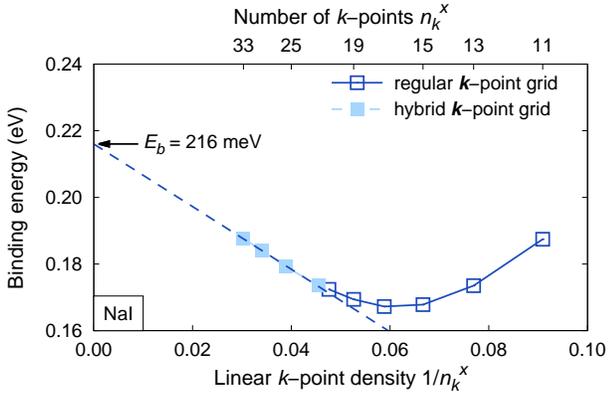}
  \caption{
    Convergence of binding energy of lowest exciton state for NaI. 
  }
  \label{fig:bse_convergence_NaI}
\end{figure}

As mentioned in \sect{sect:computational_parameters} the slow convergence of exciton binding energies with $\kb$-point sampling requires an extrapolation scheme \cite{FucRodSch08} the applicability of which is contingent upon reaching the linear regime. We illustrate this approach for NaI in \fig{fig:bse_convergence_NaI}, where it can be seen that linear behavior is not quite accomplished for regular meshes containing as many as $21\,\times\,21\,\times\,21$ $\kb$-points. However, denser sampling was possible by using hybrid $\kb$-point meshes \cite{FucRodSch08} that are well suited in the case of a material with approximately parabolic bands. This denser sampling justifies a linear fit of the exciton binding energy (see \fig{fig:bse_convergence_NaI}) and extrapolation yields a value for the lowest excitonic bound state of 216\,meV. This number is in good agreement with experimental measurements at 80\,K, which yield a binding energy of 240\,meV \cite{EbyTeeDut59}.

\begin{figure}
\includegraphics[scale=\myscale]{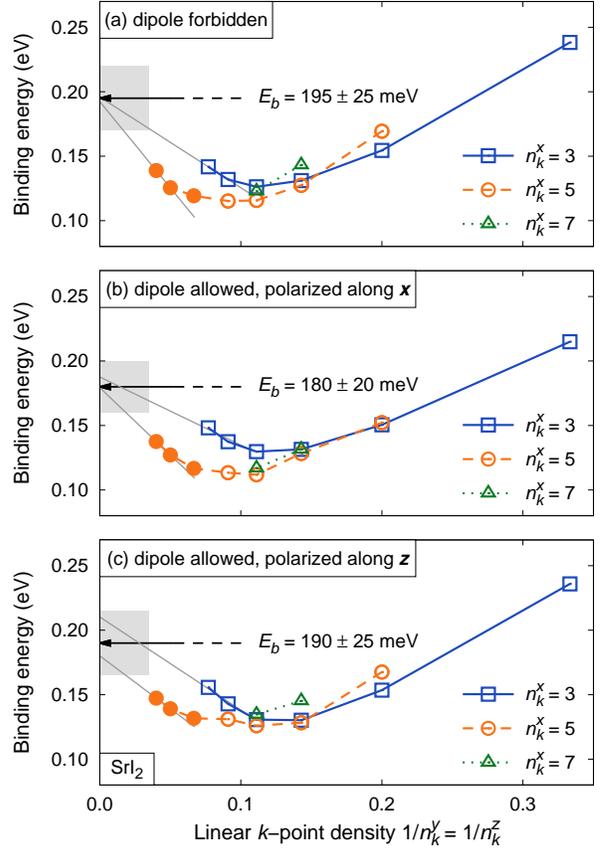}
  \caption{
    Convergence of exciton binding energies in SrI$_2$ from BSE with respect to $\kb$-point sampling for (a) dipole forbidden and (b,c) dipole allowed transitions. Empty and filled symbols denote binding energies calculated using standard and hybrid meshes, respectively. Arrows and gray bars indicate extrapolated binding energies and associated error estimates.
  }
  \label{fig:bse_convergence_SrI}
\end{figure}

As illustrated in \fig{fig:SrI2_gw}(d), the band structure of SrI$_2$ is much more complicated than the one for NaI. The minimum of the lowest conduction band is less pronounced and, as a consequence, a large number of very flat conduction bands lie within a small energy range and are likely to contribute to the lowest excitonic states. This situation is further exacerbated by the large number of very shallow valence bands with several extrema that are energetically close [see \fig{fig:SrI2_gw}(d)] and, hence, are also expected to contribute to the formation of the lowest excitonic states. In addition, unlike NaI, SrI$_2$ is not of cubic symmetry, hence, the exciton localization is generally anisotropic, an aspect that will be explored in more detail below. This situation motivates the application of anisotropic $\kb$-point meshes when converging exciton binding energies using both regular (empty symbols in \fig{fig:bse_convergence_SrI}) and hybrid grids (filled symbols in \fig{fig:bse_convergence_SrI}).
 
Due to the symmetry of the states that are involved, the lowest bound exciton state corresponds to an optically (dipole) forbidden transition. In \fig{fig:bse_convergence_SrI} we show the dependence of the binding energy on $\kb$-point sampling for this state as well as for the first states that are dipole allowed for $x$-polarized ($\vec{E}\,||\,\vec{x}$) and $\vec{z}$-polarized ($\vec{E}\,||\,\vec{z}$) light, respectively. The most highly converged data point in \fig{fig:bse_convergence_SrI} corresponds to 685 $\kb$-points in the full BZ and requires setting up and (iteratively) diagonalizing a BSE matrix with a rank of over 140,000. We extract binding energies by linear extrapolation \cite{FucRodSch08} based on the two most highly converged data points for the data sets corresponding to $n_k^x=5$ and $n_k^x=7$ as indicated in \fig{fig:bse_convergence_SrI}.

In this fashion we obtain a binding energy of $195\pm25\,\meV$ for the dipole forbidden state from separate fits to regular and hybrid meshes, where the error estimate is based on the deviation between the fits. For the dipole allowed excitons we obtain binding energies of $180\pm20\,\meV$ ($\vec{x}$-polarized) and $190\pm25\,\meV$ ($\vec{z}$-polarized). These values are in rough agreement with the estimate of 260\,meV obtained by Pankratov \etal\ based on the effective mass approximation \cite{PanPopShi13}. Note, however, that this agreement should be considered rather fortuitous due to the strong anisotropy of the hole effective mass tensor. The analysis of the excitonic wave functions in the following section will provide further evidence for anisotropic excitonic properties.

\subsection{Exciton densities and electron-hole separation}
\label{sect:excdens}

\begin{figure*}
\includegraphics[scale=\myscale]{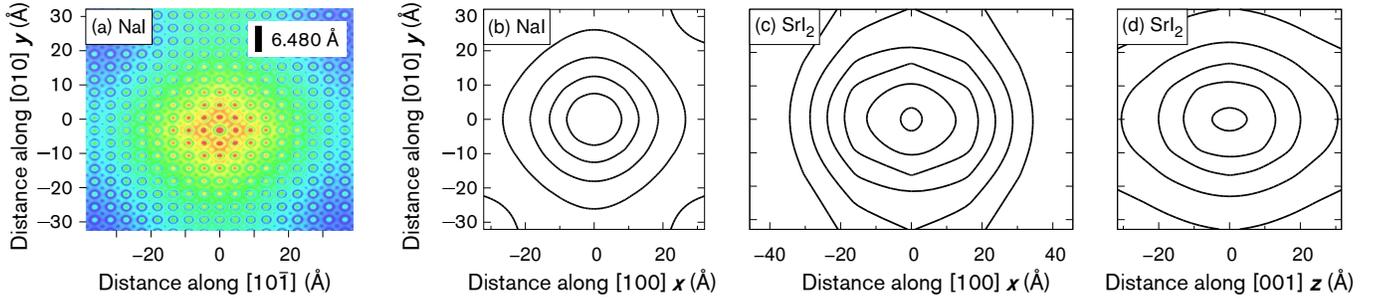}
  \caption{
    (a) The logarithm of the electron-hole density $\rho_{eh}(\rb_e, \rb_h)$ defined in \eq{eq:excden} for the exciton ground state state of NaI in a $\left\{110\right\}$ plane assuming the hole is located at an iodine site, i.e. $\rb_h=0$. The black bar indicates the lattice constant. Also shown are logarithmic contour maps of the envelope function $\widetilde{g}_{eh}(\rb)$ defined in \eq{eq:envelope} for (b) the ground state exciton of NaI and (c,d) the lowest lying dipole forbidden exciton state of SrI$_2$ in projected onto (001) and (100).
  }
  \label{fig:envelopes}
\end{figure*}

\begin{figure}
  \centering
\includegraphics[scale=\myscale]{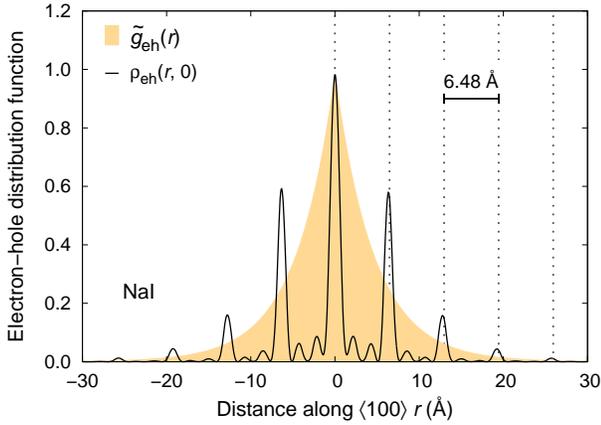}
  \caption{
    Envelope function $\widetilde{g}_{eh}(r)$ and exciton electron density $\rho_{eh}(r,0)$ for NaI as defined in Eqs.~(\ref{eq:envelope}) and (\ref{eq:excden}) projected onto $\left<100\right>$ direction with the hole placed on an iodine ($\rb_h+\Rb_h=0$). The vertical dotted lines indicate the lattice spacing ($a_0=6.480\,\AAA$).
  }
  \label{fig:env_vs_den} 
\end{figure}

The cell-periodic electron and hole densities [see Eqs.~(\ref{eq:rhoe}) and (\ref{eq:rhoh})] for the three degenerate exciton ground state states in NaI are spherically centered around iodine atoms, reflecting the fact that the heavy hole bands are almost exclusively formed from iodine $p$-states. The conduction bands of NaI are strongly hybridized and thus it is not entirely obvious that also the electron will localize on iodine atoms. This picture is, however, corroborated by visualizing the electron charge density of the exciton with the hole placed at an iodine atom as shown in \fig{fig:envelopes}(a). The exciton ground state envelope function shown in \fig{fig:envelopes}(a) is spherically symmetric as well. The deviations from spherical symmetry at large electron-hole separation that are apparent in \fig{fig:envelopes}(a) are an artifact of finite $\kb$-point sampling and the periodic boundary conditions applied in our calculations. Fitting the envelope function to the charge density of a $1s$ hydrogen-like wave function $\rho(\rb)\propto\exp[-2r/a]$ yields a Bohr radius (defined as the most probable electron-hole distance) of $a=9.3\,\text{\AA}$ for a $14^3$ $\Gamma$-centered MP grid. This corresponds to an exciton binding energy of 210\,meV in very good agreement with the extrapolated value of 216\,meV obtained above. Thus it is no surprise that for NaI also the effective mass approximation is reasonably accurate yielding a binding energy of 256\,meV.

Figure~\ref{fig:env_vs_den} illustrates the relation between the envelope function $\widetilde{g}_{eh}(\rb)$ introduced in Eqs.~(\ref{eq:envelope}--\ref{eq:eh_pair_distribution}) and the electron-hole density $\rho_{eh}(\rb_e,\rb_h)$ in the case of NaI. While the latter exhibits fluctuations that obey the lattice periodicity, the envelope function is smooth and decays monotonically and exponentially. The lattice periodicity enters in \eq{eq:eh_pair_distribution} via the excitonic single-particle densities $\rho_e(\rb_e)$ and $\rho_h(\rb_h)$.

\begin{figure}
  \centering
\includegraphics[width=\columnwidth]{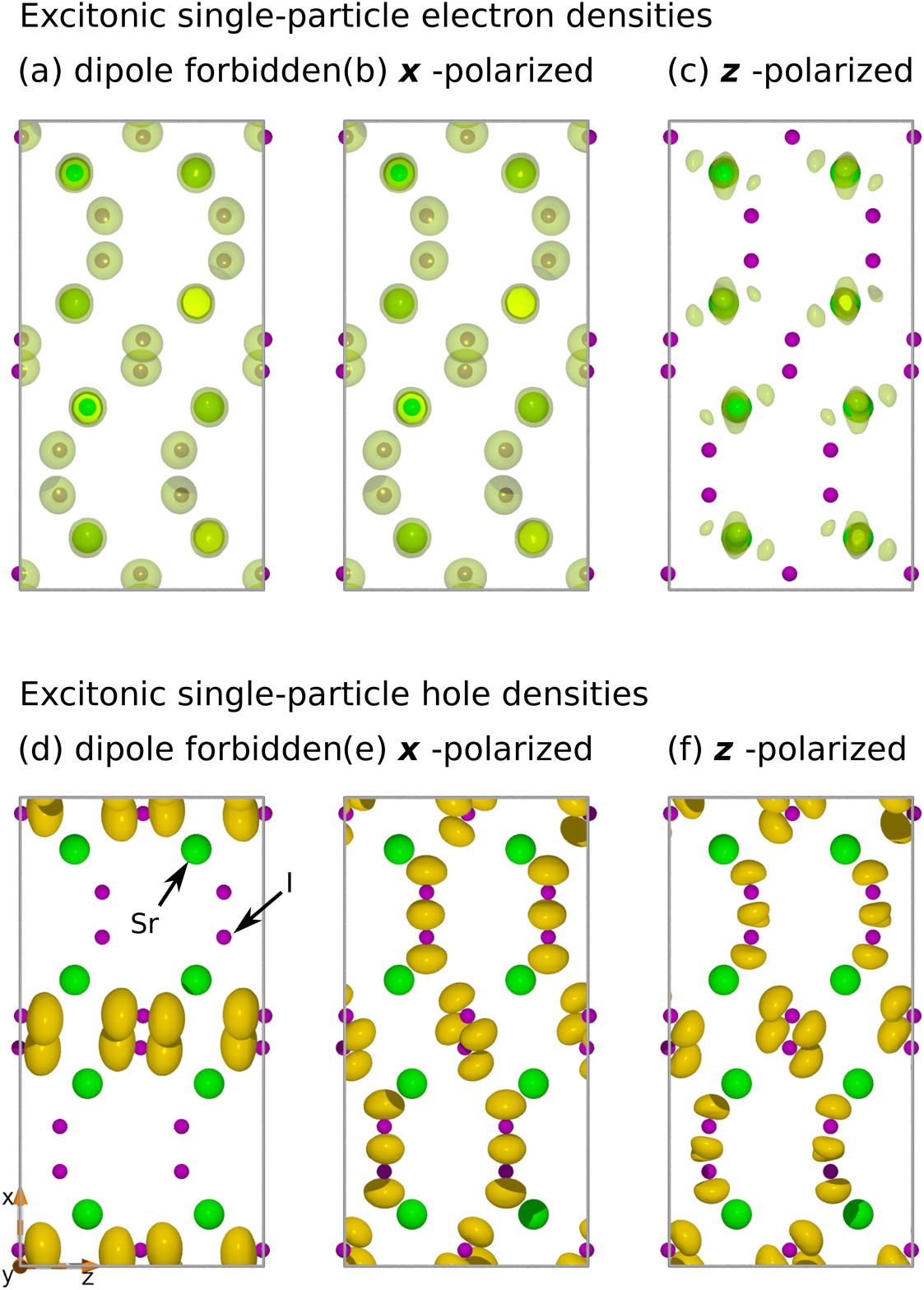}
  \caption{
    Excitonic single-particle (a-c) electron $\rho_e(\rb_e)$ and (d-f) hole densities $\rho_h(\rb_h)$ according to Eqs.~(\ref{eq:rhoe}) and (\ref{eq:rhoh}) for the lowest energy dipole forbidden, $\vec{x}$-polarized, and $\vec{z}$-polarized excitons in SrI$_2$. Small purple and large green spheres indicate I and Sr ions, respectively.
  }
  \label{fig:sri_dens}
\end{figure}

SrI$_2$ again proves itself much more complex and appears to be not well suited for a simple description in terms of the effective mass equation. The hole and electron densities of the three states discussed in the previous section are displayed in \fig{fig:sri_dens}. In the dipole forbidden and $\vec{x}$-polarized states the electron densities exhibit $s$-like maxima centered on all atoms with equal amplitude, whereas the Sr\,$4d$-character dominates in the $\vec{z}$-polarized case. Note that the exciton single particle densities shown in \fig{fig:sri_dens}(a-c) does not simply coincide with the single-particle density of the lowest lying conduction band state at $\Gamma$, which primarily exhibits Sr\,$5s$-character.

The hole densities shown in \fig{fig:sri_dens}(d-f) display I\,$p$-character for all three considered states. It is noteworthy that the dipole forbidden state only has appreciable hole density around iodine atoms belonging to the I(2) set of Wyckoff sites (compare \sect{sect:computational_parameters}) as shown in \fig{fig:sri_dens}(d). The same behavior is observed for other dipole forbidden states. In the dipole allowed excitons we find hole densities of equal amplitude on all iodide atoms.

The envelope functions for the SrI$_2$ excitons discussed here differ from those of NaI in several aspects:
({\em i}) They do not display hydrogenic ($\rho(\rb)\propto\exp[-2r/a]$) density dependence, at least not in the vicinity of the center-of-mass. Rather, by fitting several different functional types, an ellipsoidal Gaussian function was judged to most closely reproduce the BSE result. In principle, it might be possible to benchmark our results against the eigenvalues of the anisotropic effective mass equation \cite{Dev69, Sch97}. This is, however, beyond the scope of the present work.
({\em ii}) Because of the crystal symmetry the envelope functions in SrI$_2$ are anisotropic as illustrated in \fig{fig:envelopes}(c,d). For the dipole forbidden state the best fit resulted in anisotropic full width at half maxima (FWHM) of 21, 17, and 21\,\AA\ in the $x$, $y$, and $z$-directions, respectively.
({\em iii}) The effective Bohr radius of 9.3\,\AA\ for NaI corresponds to a FWHM of 6.4\,\AA. As a result, despite the apparent dominance of the localized Sr\,$4d$ states of the conduction bands, the lowest lying SrI$_2$ excitons have a spread which is about two to three times larger than the $1s$ exciton in NaI. This is solely due to the single $s$-like minimum at the $\Gamma$-point [see top panel of \fig{fig:SrI2_gw}(d)].

\section{Summary and Conclusion}
\label{sect:discussion}

In this work we have studied from first principles, single and two-particle excitations in two prototypical scintillator materials, NaI and SrI$_2$. This was motivated by the need to understand the role of free excitons in scintillator non-proportionality. The two systems were judiciously chosen to represent two significantly different responses to incident radiation. NaI is a standard scintillator material with strongly non-linear dependence of light-yield as a function of incoming photon/electron energy, while SrI$_2$ has recently been discovered to have excellent proportionality.

On the basis of DFT and $G_0W_0$ calculations we obtained rather similar band gaps of 5.5 and 5.2\,eV for NaI and SrI$_2$, respectively. The dielectric functions of NaI and SrI$_2$ displayed significant red shift of the oscillator strengths due to excitonic effects. As a result the optical spectra calculated from BSE deviate substantially both in intensity and structure from the single-particle RPA spectra calculated with DFT+$\Delta$ over the energy range considered in this study, i.e. up to approximately 6\,eV above the conduction band edge. Although the difference is expected to diminish at high energies, these results highlight the need for incorporating excitonic effects in the dielectric models used in the study of carrier and exciton generation by the photoelectrons during the cascade. 

Almost all models for scintillation require knowledge of the binding energy of excitons, which determines their population relative to free carriers at very early times $<1\,\text{ps}$ after the impact of the ionizing radiation. In this regard the main result of this work is that the calculated groundstate exciton binding energies fall in the range between 200 and 220\,meV for both NaI and SrI$_2$. Hence, the superior energy resolution of SrI$_2$ cannot be directly correlated with the population of the free excitons. 

To study localization and spatial structure of excitons we introduced a decomposition of the six-dimensional exciton wave function int a product of single-particle densities and an envelope function. In the case of NaI we obtained a perfectly spherical envelope function, which can be fit very well assuming a hydrogen-like wave function. The binding energy obtained in this way is in good agreement with the values obtained directly from BSE calculations and via the effective mass approximation, which is expected given the hydrogen-like character of the excitonic groundstate.

The SrI$_2$ low-energy excitons revealed more complex character, which could not be fitted using simple hydrogenic wave functions. The envelope function was found to possess anisotropic first moments in accord with its anisotropic effective mass tensor. More surprising is the finding that excitonic wave functions in SrI$_2$ are more extended than in NaI, which suggests that non-radiative exciton-exciton annihilation mediated by exchange is stronger in SrI$_2$. This is unexpected as such processes contribute to stronger non-proportional response in conventional models of scintillation. This study thus calls for more detailed investigation of the temporal evolution of carrier density during early stages of scintillation from first principles, and considering the relative importance of free versus self-trapped excitons in scintillator non-proportionality.

\begin{acknowledgments}
We acknowledge fruitful discussions with C.~R\"odl. This work was performed under the auspices of the U.S. Department of Energy by Lawrence Livermore National Laboratory under Contract DE-AC52-07NA27344 with support from the National Nuclear Security Administration Office of Nonproliferation Research and Development (NA-22). P.E. acknowledges support through the ``Areas of Advance -- Materials Science'' at Chalmers and computer time allocations by the Swedish National Infrastructure for Computing at NSC (Link\"oping) and C3SE (Gothenburg).
\end{acknowledgments}

\end{document}